\documentclass[sigconf]{acmart}

\AtBeginDocument{%
  \providecommand\BibTeX{{%
    \normalfont B\kern-0.5em{\scshape i\kern-0.25em b}\kern-0.8em\TeX}}}

\setcopyright{acmcopyright}
\copyrightyear{2022}
\acmYear{2022}
\acmDOI{10.1145/3546096.3546110}

\acmConference[CSET 2022]{Cyber Security Experimentation and Test Workshop}{August 8, 2022}{Virtual, CA, USA}
\acmBooktitle{Cyber Security Experimentation and Test Workshop (CSET 2022), August 8, 2022, Virtual, CA, USA} 
\acmPrice{15.00}
\acmISBN{978-1-4503-9684-4/22/08}

\usepackage{xspace}

\newcommand{\eg}{\textit{e.g.,~}}
\newcommand{\cf}{\textit{cf.,~}}
\newcommand{\ie}{\textit{i.e.,~}}
\newcommand{\etal}{\textit{et al.}\xspace}

\begin{document}

\title{Position Paper: Certificate Root Stores---An Area of Unity or Disparity?}

\author{Jegan Purushothaman}
\affiliation{%
  \institution{Carleton University}
  \city{Ottawa}
  \country{Canada}
}

\author{Ethan Thompson}
\affiliation{%
 \institution{Carleton University}
 \city{Ottawa}
 \country{Canada}
}

\author{AbdelRahman Abdou}
\affiliation{%
  \institution{Carleton University}
  \city{Ottawa}
  \country{Canada}
}


\begin{abstract}
Organizations like Apple, Microsoft, Mozilla and Google maintain certificate root stores, which are used as trust anchors by their popular software platforms. Is there sufficient consensus on their root-store inclusion and trust policies? We measure disparities among their root stores, accounting for various aspects such as inclusion policies, delivery methods, trust context, and the certificates themselves. Disparities appear astounding, including in the government-owned certificates that they trust. We believe such a status-quo is alarming, and warrants more attention from the wider community.

\end{abstract}




\maketitle

\section{Introduction}
A web browser must obtain a web server's cryptographic public key to verify the identify of that server. To prevent an intercepting attacker from substituting the server's key with their own, the server sends the browser a \emph{certificate}---a document binding the server's domain name to a public key. This document is digitally signed and issued by a third party, called a Certificate Authority (CA). For a browser to validate the document's signature, the CA's own public key must be accessible to the browser. This key is included in another certificate, which is either stored locally on the machine that the browser is running on, or is sent along with the server's certificate. Browsers trust locally-stored certificates, so such certificates need not be signed by another CA; they are thus signed by the same CA that owns the enclosed public key, \ie \emph{a self-signed certificate}. But when a CA is not trusted, its certificate must be sent along with the server's, and it needs to be validated in the same manner as the server's certificate. Instead of validating one certificate, the browser thus validates a chain of certificates. Regardless of the chain's length, it must lead to a locally-stored (\ie client-trusted) certificate, or else the browser fails to verify the server's identity. The collection of the locally-stored certificates is called the \emph{root store}. They constitute the anchor of trust. 

The above process is not exclusive to web browsers. It applies to any network application that relies on certificates for establishing secure sessions with a verified server, including secure email (when a local mail client is used, such as Apple's Mail application for iOS or Thunderbird), remote logins (like SSH), and more recently the Domain Name System (DNS)~\cite{JahromiDoTcert21}. The standard format for such certificates is called X.509, which is standardized by the International Telecommunications Union (ITU). Application builders have a choice: either rely on the root store of the Operating System (OS) upon-which the application will run, or build and maintain their own root store. In the latter, the root store is typically part of the application's software package, downloaded when the application is first installed. Mozilla has long maintained its own root store, and the organization's products, famously the Firefox browser, rely on this store. Apple Safari and Microsoft Edge rely on the OS's root store. Google runs a root store for Android. Its Chrome browser had been relying on the OS's root store; it recently got its own root store.\footnote{\url{https://www.chromium.org/Home/chromium-security/root-ca-policy}}

One quickly realizes how critical root stores are to an ecosystem. A malicious CA whose certificate is in the root store can, in theory, substitute \emph{any} server's public key (not just a server that has asked this CA for a certificate), thus impersonate \emph{any} website~\cite{chuat2020sok}. Root store management is thus of extreme importance. For example, unused/untrusted certificates should be removed~\cite{perl2014you} to reduce the trust base---\cf the \textsc{Small-Trusted-Bases} security principle~\cite{van2020computer}. In practice, each vendor maintaining a root store has its own standards to determine policies like who to trust (\ie which organization will have its certificate included in the root store), what applications will a root certificate be trusted for, how the root store is communicated to clients, and when to distrust an organization. Ideally, a user would understand the effect of their decision, \eg to use Chrome on Linux or Firefox on Windows, on their realm of trust. But this is unlikely to be the case in practice. 

Root store management policies are influenced by a variety of factors, like ethical principles, politics, and business considerations, which have resulted in increasingly widening deltas between popular root stores in the Public Key Infrastructure (PKI) ecosystem. Root store sizes were compared in 2015~\cite{fadai_2015_trust}, where it was found that back then their sizes already varied widely.

Beyond the differences in root store sizes, a few question arise. Taking a current snapshot in time, to what degree is the variation among the trusted certificates themselves? Do such differences divide the Internet into distinct trust zones? Can a user's choice to use one platform over another substantially change the realm they (implicitly) trust? 
We shed some light on these questions, focusing on five major root stores maintained by five vendors: Apple (MacOS/iOS), Microsoft (Windows OS), Mozilla (Firefox/Thunderbird), Canonical (\emph{Ubuntu} henceforth), and Google (Android). Recent research confirms that Apple, Microsoft, and Mozilla constitute the most popular root stores~\cite{ma2021tracing}. We will refer to Google's root store as \emph{Android} from this point on, to avoid confusion with Google's recent Chrome Root Program.

\section{Disparities Between Vendors}
We breakdown this section by the four stages in the life of a potential root-store certificate (or \textit{root certificate} for short). Specifically, we discuss: (1) root-store inclusion policies, (2) methods of delivering root stores to clients, (3) the root certificates themselves, and (4) root certificate usage across applications.

\subsection{Differences in policies of inclusion}
A noticeable difference exists in the level of details provided by each vendor on their root-store inclusion policy. Mozilla's process is relatively transparent, where discussions pertaining to a root certificate's inclusion occurs on a publicly accessible forum.\footnote{\url{https://groups.google.com/g/mozilla.dev.security.policy} and \url{https://wiki.mozilla.org/CA}} Microsoft has made available a significant amount of information detailing the application process and requirements for a CA to join the Microsoft Trusted Root Program,\footnote{\url{https://docs.microsoft.com/en-us/security/trusted-root/program-requirements}} but final inclusion decisions are ultimately under the company's control. Apple makes available relatively the least amount of information on root-store inclusion policies.\footnote{\url{https://www.apple.com/certificateauthority/ca_program.html}} Not much information is provided, beyond listing a few brief criteria that an organization has to satisfy for including its certificate into Apple's root store.

The root stores of Ubuntu, Android, and many other open-source platforms are primarily based on Mozilla's, which allows them to push the administrative burden to Mozilla.\footnote{\url{https://answers.launchpad.net/ubuntu/+source/ca-certificates/+question/693748}} 
To include a root certificate in Ubuntu and Android, firstly a CA must add their certificate to the Mozilla Root Store, upon which they contact the respective vendor requesting an inclusion and provide details about their root certificate inclusion by Mozilla.

Apple, Microsoft, Mozilla, and Google are part of the CA/Browser (CAB) Forum, as a result much of the technical details pertaining to the certificate and administrative regulations of a CA's behavior have been standardized among the four. A common aspect for root-store inclusion among them is the requirement for the CA to be audited by an independent and qualified authority. CAs are typically expected to complete a \emph{WebTrust Principles and Criteria for Certification Authorities} audit, or an equivalent from the European Telecommunications Standards Institute (ETSI). These audits must be run by accredited organizations, which include large accounting firms like Ernst\&Young (EY) and PriceWaterhouseCoopers (PwC).\footnote{A list of Webtrust practitioners can be found here: \url{https://www.cpacanada.ca/en/business-and-accounting-resources/audit-and-assurance/overview-of-webtrust-services/licensed-webtrust-practitioners-international}} A CA can also provide audits of equivalent standards, however it is unclear exactly under what circumstances would this be accepted over a WebTrust/ETSI audit, or how an equivalent audit may influence restrictions on a commercial CA. There are different types of WebTrust/ETSI audits available, and all vendors appear to be generally in agreement on which audits need to be conducted. 

Interesting differences were noticed between Microsoft and the others when we look at the inclusion of government root certificates. Of the five vendors, Microsoft appears to be the only one explicitly differentiating between a government root certificate and a commercial one (\ie CA), with the option for a government to have their certificates \emph{domain restricted} if they did not submit an audit that meets the WebTrust/ETSI standard. Microsoft states:
\begin{center}
\emph{``Government CAs must restrict server authentication to government-issued top level domains and may only issue other certificates to the ISO3166 country codes that the country has sovereign control over [..]. These government-issued TLDs are referred to in each CA's respective contract.''}
\end{center}
It appears that such a contract, between a government and Microsoft, is intended to legally bind the government to issue certificates only to its domains. The \textsc{Name Constraint} field of an X.509 certificate is designed for such a purpose, but it does not apply to self-issued certificates (typically root certificates)~\cite[p.40]{rfc5280}, and so it is unclear if this part of the contract is technically enforced.

Mozilla solicits the broader community's input on concerns about a root certificate inclusion. A government's record on areas such as human rights, domestic surveillance, freedom of speech, and foreign policy can impact its likelihood of root-store inclusion. For Microsoft, governments can be included in the root store with the understanding that their certificate will be domain-restricted. This contributes to the differences in the number of government root certificates, as we show below.

\subsection{Root Store Delivery}
Unlike the root stores of Android, Apple, and Ubuntu, which are fully included in the OS upon installation, Microsoft's is not fully included on a fresh installation of Windows~\cite{fadai_2015_trust}. Since Windows Vista, each Windows installation contains only a subset of Microsoft's root store. The full store is available in an online repository, which is accessed as needed by applications through OS-update avenues, thus providing Windows with the capability of expanding its local root store on the fly. Mozilla's root store is contained within the Network Security Services (NSS) libraries, which support Mozilla's Transport Layer Security (TLS) and other security standards. NSS is used across Mozilla's products, including Firefox and Thunderbird.

\subsection{Root Store Differences}
An X.509 certificate has a unique serial number, included in the \textsc{Serial Number} field. The number is unique per organization. Accordingly, a certificate can be uniquely identified (world-wide) by combining the \textsc{Issuer} field with \textsc{Serial Number}.

\begin{figure}[htbp]
\centerline{\includegraphics[width = 0.5 \textwidth]{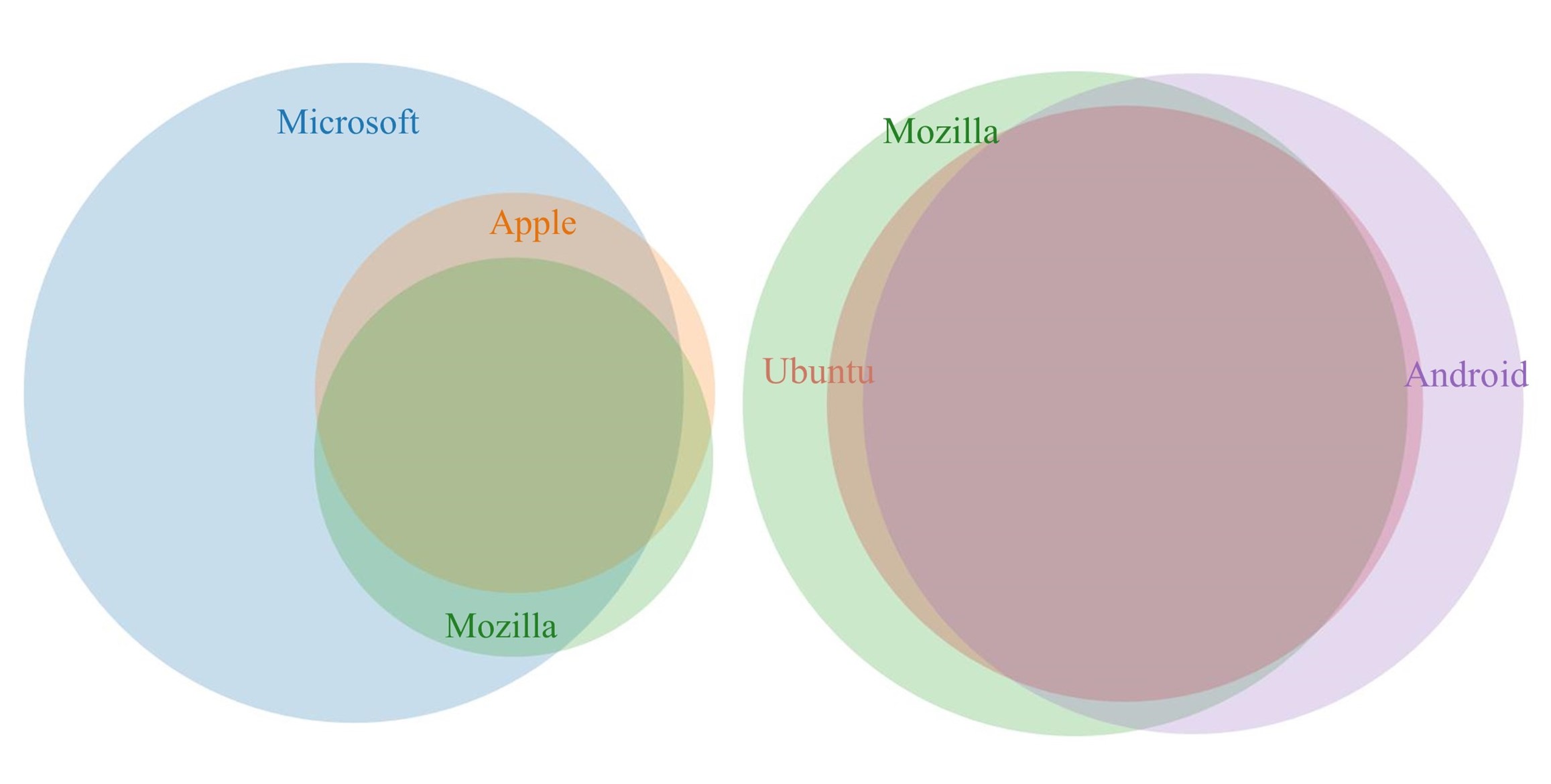}}
\caption{Root Certificate Overlap}
\label{fig:certoverlap}
\end{figure}

We use Venn Diagrams to visualize the degree of overlap between root stores. Because Android's and Ubuntu's root stores are slight variations of Mozilla's, we plot two Venn Diagrams: one for Microsoft, Mozilla, and Apple, and the other for Mozilla, Android and Ubuntu. Figure~\ref{fig:certoverlap} shows the two diagrams, drawn on different scales.

Microsoft's root store is the largest. It contains 432 certificates, 247 of which are present only in Microsoft's root store. Mozilla's root store  (green) is the smallest of all three, with only 158 certificates, and has the highest degree of overlap with the other two. Almost 95\% of Mozilla's trusted certificates are also trusted by one (or both) of the other two vendors. Mozilla's root store is almost fully contained within Microsoft's with 145 certificates overlapping. It has ten certificates that are neither trusted by Apple nor Microsoft. Apple's root store (orange) size sits close to Mozilla's.. It has 159 certificates, 7 of which are trusted only by Apple; 94\% of Apple's root store overlaps with Microsoft's, but only 78\% overlaps with Mozilla's. The three sets intersect in 123 certificates, which is 24.5\% of their union (503 certificates).

For the overlap between NSS-based root stores, namely Mozilla, Ubuntu, and Android (the diagram to the right of Fig.~\ref{fig:certoverlap}), Android's root store (purple) contains about as many certificates as Mozilla's. It has 156 certificates, 11 of which are exclusive to Android's (\ie only trusted by Android). Mozilla's root store contains the second largest number of exclusive certificates: 10. Ubuntu's root store has a higher overlap with Mozilla's. Ubuntu has the smallest root store---127 certificates, and there are no certificates exclusively trusted by Ubuntu. The slight differences between the three sets make it clear that being based on NSS does not necessarily mean an exact replica. Occasionally, each vendor creates exceptions for some certificates that it exclusively trusts (or distrusts). Note that each such occasional exception potentially affects the trust realm of millions of users world-wide.

\begin{figure}[htbp]
\centerline{\includegraphics[width = 0.5 \textwidth]{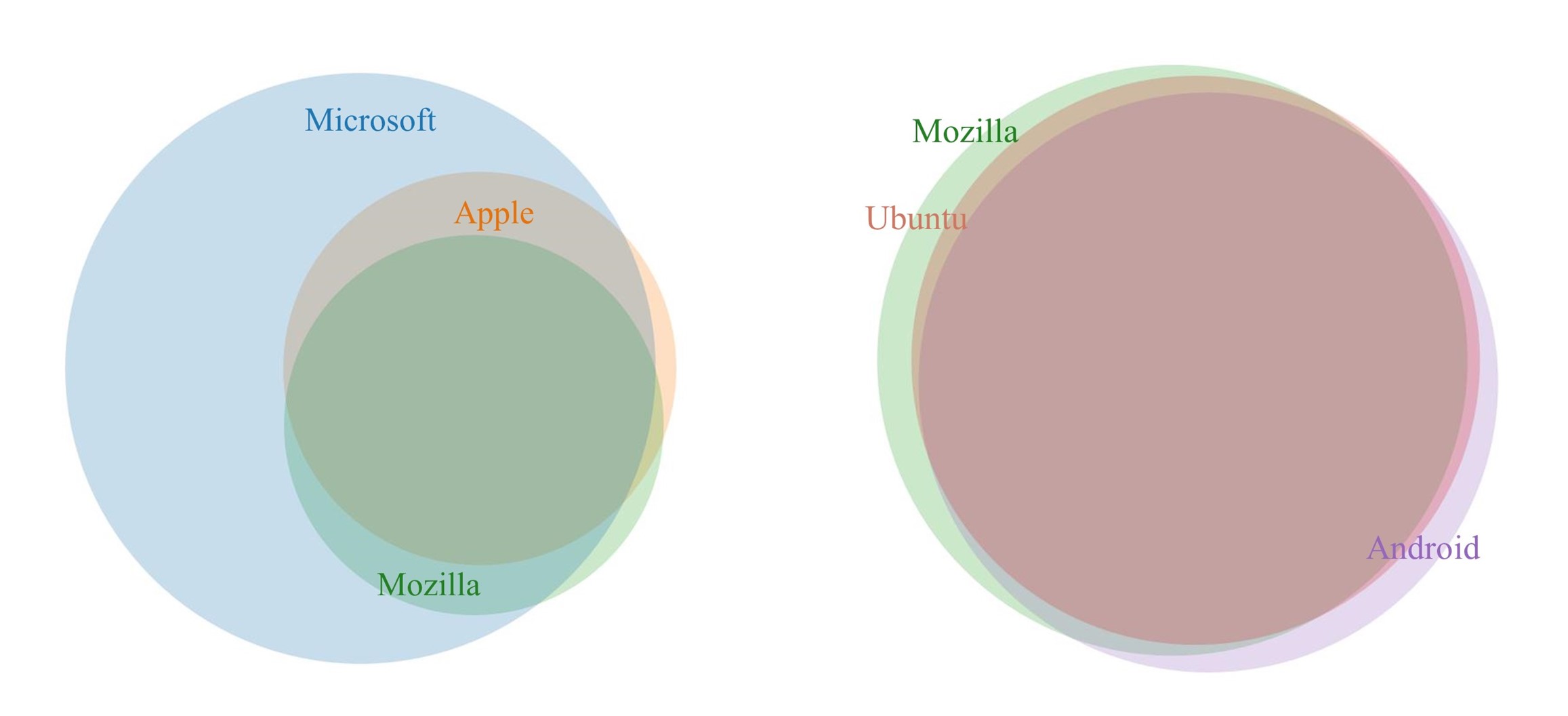}}
\caption{Root Organization Overlap}
\label{fig:CAorgoverlap}
\end{figure}

To investigate overlaps in trusted organizations, we look at the organizations that issued the root certificates in each root store. This also eliminates alternate, expired, or newer versions of the same root certificate, which would be counted as exclusive in Fig.~\ref{fig:certoverlap}.

Figure~\ref{fig:CAorgoverlap} contains Venn Diagrams showing organization overlaps between the root stores. Again, Microsoft dominates over the three major root stores when it comes to the organizations represented. Microsoft's root certificates are issued by 133 different organizations, 69 of which are exclusively represented on Microsoft's root store. Many of these are government organizations. Mozilla's root certificates are issued by 55 organizations, two of which are exclusively trusted by Mozilla. Apple once again sits in the middle, trusting 59 different organizations, none of which are exclusively trusted by Apple.

\subsection{Differences in trust context}
The existence of a certificate in a root store does not automatically mean it would be trusted for all applications---the trust context. While there is a \texttt{Key Usage} extension in an X.509 standard that enables the issuer to indicate how the key should be used, this is different from indicating the trust context. To elaborate, the standard defines~9 key usage options~\cite{rfc5280}:
\begin{itemize}
\item digitalSignature
\item nonRepudiation
\item keyEncipherment
\item dataEncipherment
\item keyAgreement
\item keyCertSign
\item cRLSign
\item encipherOnly
\item decipherOnly
\end{itemize}
A vendor can choose to use a CA certificate only for emails, or only for web, but this is independent from using the key as specified by the \texttt{Key Usage} extension in the certificate following the above~9 options. For specifying an application, another X.509 extension can be used: the \texttt{Extended Key Usage} extension. However, according to the standard, \emph{``this extension will appear only in end entity certificates''}~\cite{rfc5280}, and so is not typically used for root certificates.

So how do vendors implement trust context for \emph{root certificates}? This is vendor-specific. Mozilla\footnote{\url{https://udn.realityripple.com/docs/NSS/PKCS\_11\_Netscape\_Trust}} and Microsoft\footnote{\url{https://docs.microsoft.com/en-us/windows/win32/api/certenroll/nn-certenroll-ix509extensionenhancedkeyusage}} associate \emph{Trust Bits} for each certificate. Examples of trust bits include: \textsc{Server-Authentication} (for website/server/host authentication), \textsc{Code-Signing} and \textsc{Secure-Email} (for S/MIME email encryption). Even though both vendors use the same ``trust-bits'' concept, they have different such bits. For example, Mozilla has \textsc{CRL-Signing}, but Microsoft does not. Microsoft's root store supports a total of 12 trust bits---Mozilla 11. Apple, on the other hand, has 7 options:\footnote{These are not given a specific name; they are listed in the system settings as trust options given to the user.} \textsc{SSL}, \textsc{Secure-Email} (S/MIME), \textsc{Extensible Authentication} (EAP), \textsc{IPsec}, \textsc{Code Signing}, \textsc{Time Stamping}, and \textsc{X.509 Basic Policy}. It is worth noting that the standard defines six options for the \texttt{Extended Key Usage} extension.

A more urgent question is how vendors decide such trust context for root certificates, since it is not typically specified in such certificates. Unclear, but the policy is at their discretion. Apple states that it \emph{``uses a number of trust policies to determine whether a certificate is trusted. Each certificate can have a different policy''}. 

\begin{figure}[htbp]
\includegraphics[width = 0.5 \textwidth]{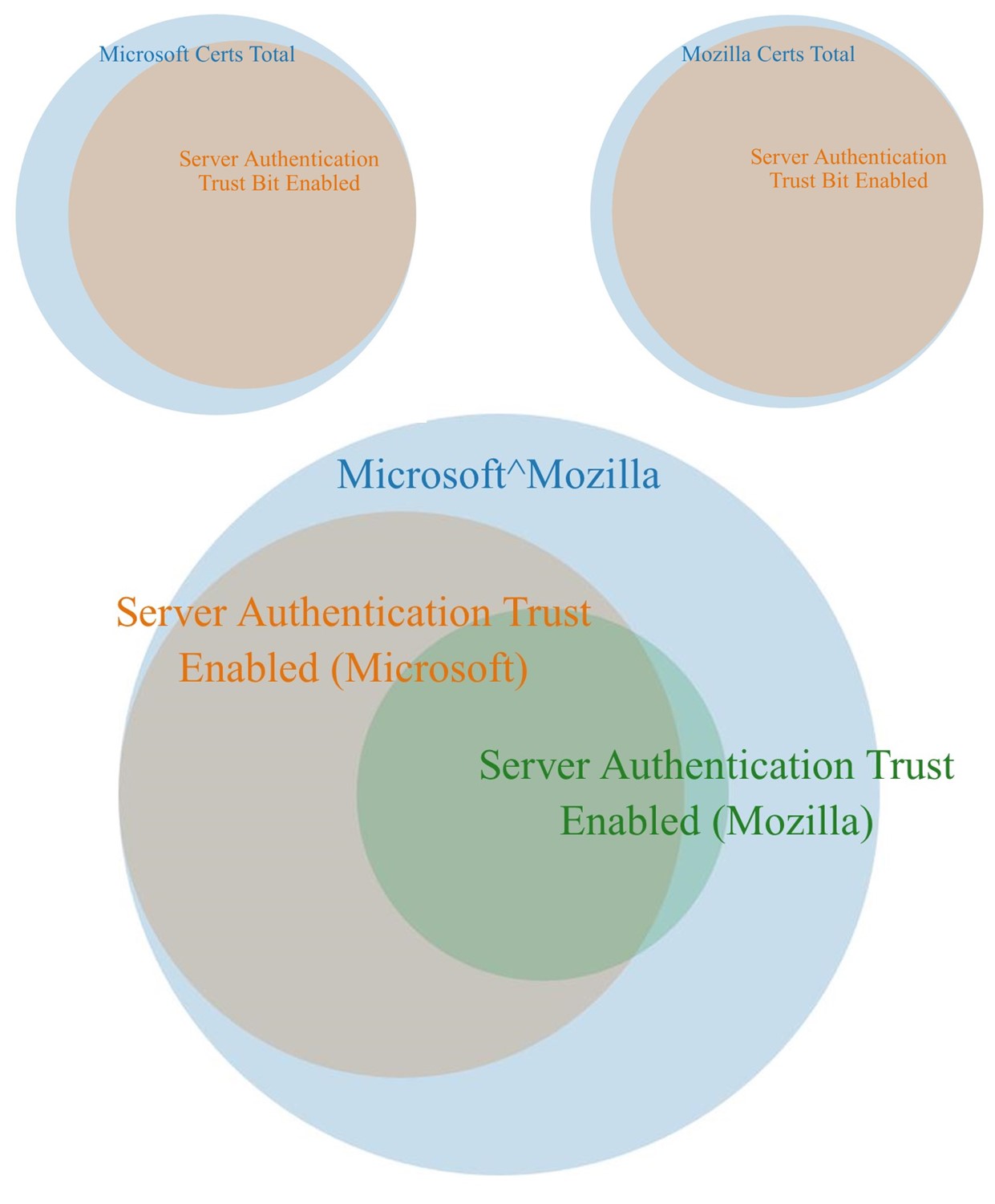}
\caption{Server Authentication Trust Bit Overlap (charts on different scales for clarity).}
\label{fig:serverauthcerts}
\end{figure}

\begin{figure}[htbp]
\centerline{\includegraphics[width = 0.5 \textwidth]{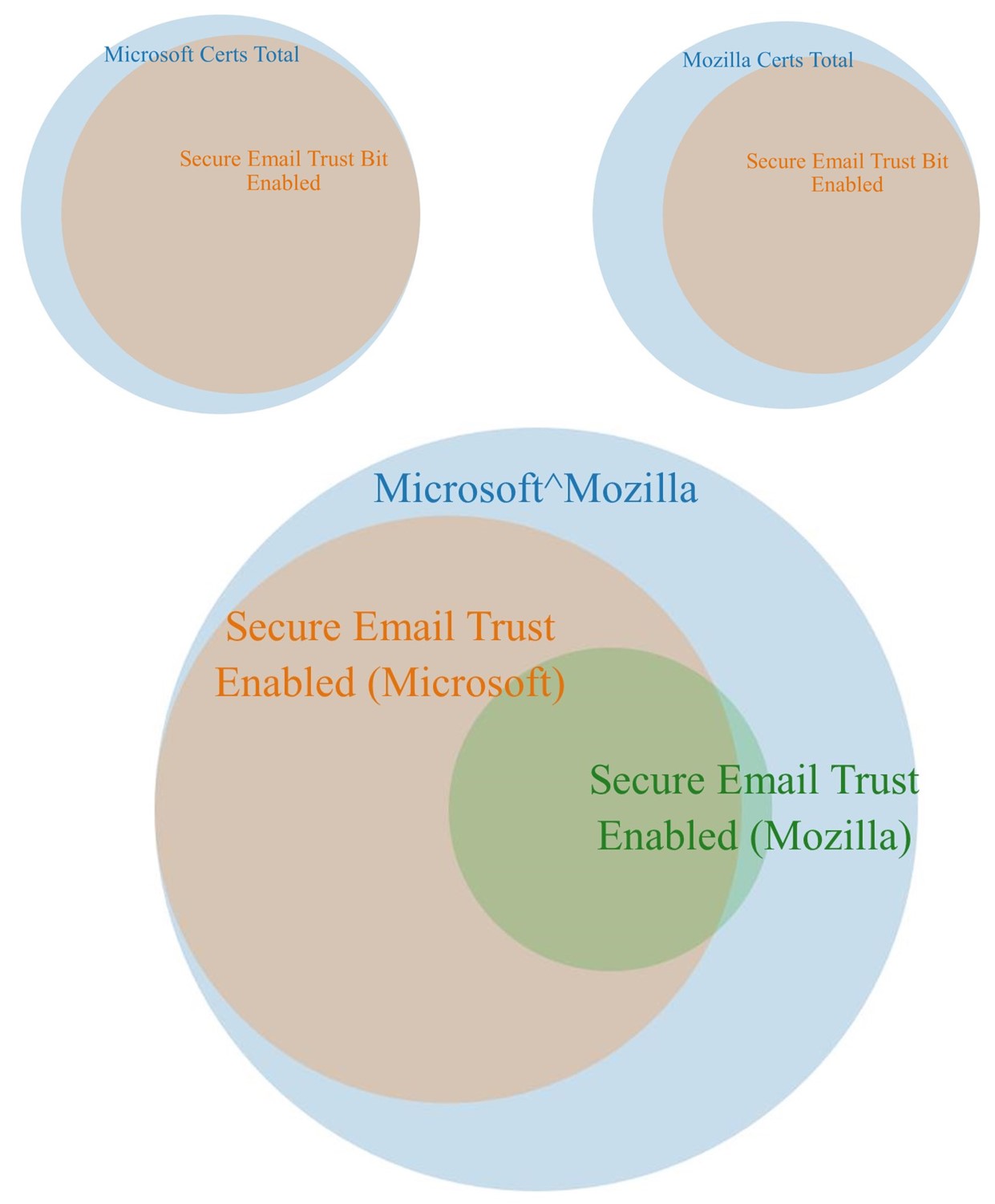}}
\caption{Secure Email Trust Bit Overlap (charts on different scales for clarity).}
\label{fig:emailcerts}
\end{figure}

We noticed that the option of modifying the trust context of a root certificate is often given to users, including on Ubuntu and Apple (MacOS). But the default trust context for each certificate appears to be yet another area of difference across vendors. For example, comparing Microsoft and Mozilla, Fig.~\ref{fig:serverauthcerts} (top) shows the proportion of root certificates trusted for server authentication on both vendors. Not all root certificates are trusted for server authentication. To measure the disparity in trust for server authentication, we create a set of certificates from the intersection of both root stores, such that each certificate in that set is trusted for server authentication by either Mozilla or Microsoft or both. The large Venn Diagram at the bottom of Fig.~\ref{fig:serverauthcerts} shows that the intersection between both sets (certificates trusted for server authentication by both vendors) is large, but not 100\%. 

Figure~\ref{fig:emailcerts} shows trust for secure email, again for Microsoft and Mozilla. A very large portion of root certificates in Microsoft's root store are trusted for secure email. The situation is slightly different for Mozilla, where only two-thirds of Mozilla's root certificates are trusted for that application. In terms of the overlap, 93\% Mozilla's root certificates that were trusted for secure email were also trusted for secure email by Microsoft. In contrast, 250 certificates were trusted by Microsoft for secure email but not by Mozilla (though 220 of these are also in Mozilla's root store).

\textbf{Summary.} The above charts show that root store overlaps is not enough to show unity between vendors; a certificate that exists in both root stores may not be trusted for the same applications.

\section{Consequences of Trust Disparities}
Do the root-store disparities noted above cause substantial disparities in trusted leaf certificates across vendors? We use Censys to approach this question. Censys\footnote{\url{https://censys.io}} collects certificates using multiple means, which include relying on CT-logs, and routinely scanning the IPv4 space. In the process, Censys validates certificate chains. It starts with the issuers of the certificates it finds, and follows the chain to determine which root store contains the chain's root (\ie which vendor would trust this certificate). Censys performs such scans and measurements on regular intervals, and makes the data available through a powerful search engine and API for security professionals and researchers. The search engine has multiple criteria for fine-tuning queries and filtering results. Currently, Censys only determines if a leaf certificate is trusted by Apple, Mozilla, and Microsoft.

We used the following example search queries to extract the data we are interested in from Censys' certificate search:
\begin{center}
\small
\texttt{tags.raw: "leaf" AND tags.raw: "trusted" AND validation.nss.valid: "true" AND NOT tags.raw: "precert"}
\end{center}
This returns currently valid leaf certificates that chain back to Mozilla's root store (trusted by Firefox). Likewise, the query:
\begin{center}
\small
\texttt{tags.raw: "leaf" AND tags.raw: "trusted" AND (validation.nss.valid: "true" AND validation.microsoft.valid:"false" AND validation.apple.valid:"false") AND (NOT tags.raw: "precert")}
\end{center}
returns currently valid leaf certificates that only chain back to Mozilla's root store (only trusted by Firefox). 

\begin{figure}[htbp]
\centerline{\includegraphics[width = 0.5\textwidth]{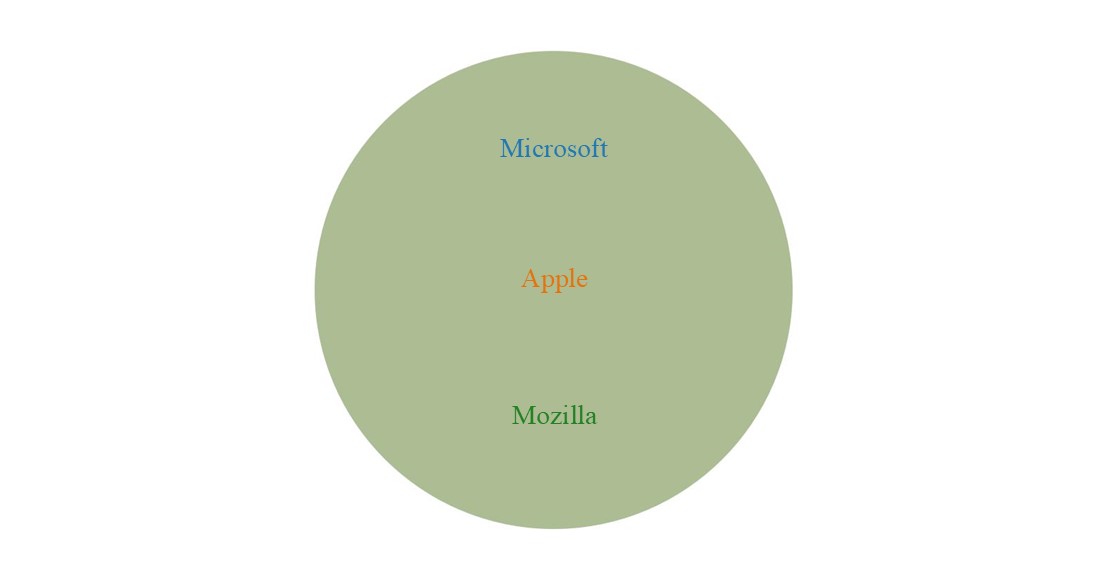}}
\caption{Leaf certificate overlap (ref: Censys)}
\label{fig:leafcert}
\end{figure}

There are 2.4 million valid (including unexpired) leaf certificates recorded on Censys as of this writing, a little over around 585 thousand of them are trusted. Almost all certificates are trusted by all three vendors (Fig.~\ref{fig:leafcert}), which means the stark root-store disparities discussed in the previous section diminish in leaf certificates. This is likely because CAs' popularity is not uniform (a small number of CAs is responsible for the majority of certificates issued worldwide). Because the overlap in trusted leaf certificates is significant, the three sets in the Venn Diagram of Fig.~\ref{fig:leafcert} appear as a single set. We thus show the number of exclusively-trusted leaf certificates by vendor in Table~\ref{tab:leafcerts}.

\begin{table}[htbp]
\caption{Trusted certificates reported by Censys.}
\begin{center}
\scalebox{1}{\begin{tabular}{c|ccc}
\toprule
& \textbf{Apple}& \textbf{Microsoft}& \textbf{Mozilla} \\
\hline
\textbf{Total certs trusted}& 585,567 & 585,595 & 584,854 \\
\textbf{Exclusively trusted}& 1 & 26 & 0 \\
\bottomrule
\end{tabular}}
\label{tab:leafcerts}
\end{center}
\end{table}

Upon investigating the 1 leaf certificate exclusively trusted by Apple, we found that it is a Developer ID certificate issued by Apple for code signing purposes. Unsurprisingly, Microsoft has the largest number of exclusively trusted leaf certificates (26). The issuers are largely among the 69 organizations that are uniquely represented on the Microsoft root store.

\section{Government Certificates}
\begin{figure}[htbp]
\centerline{\includegraphics[width = 0.5 \textwidth]{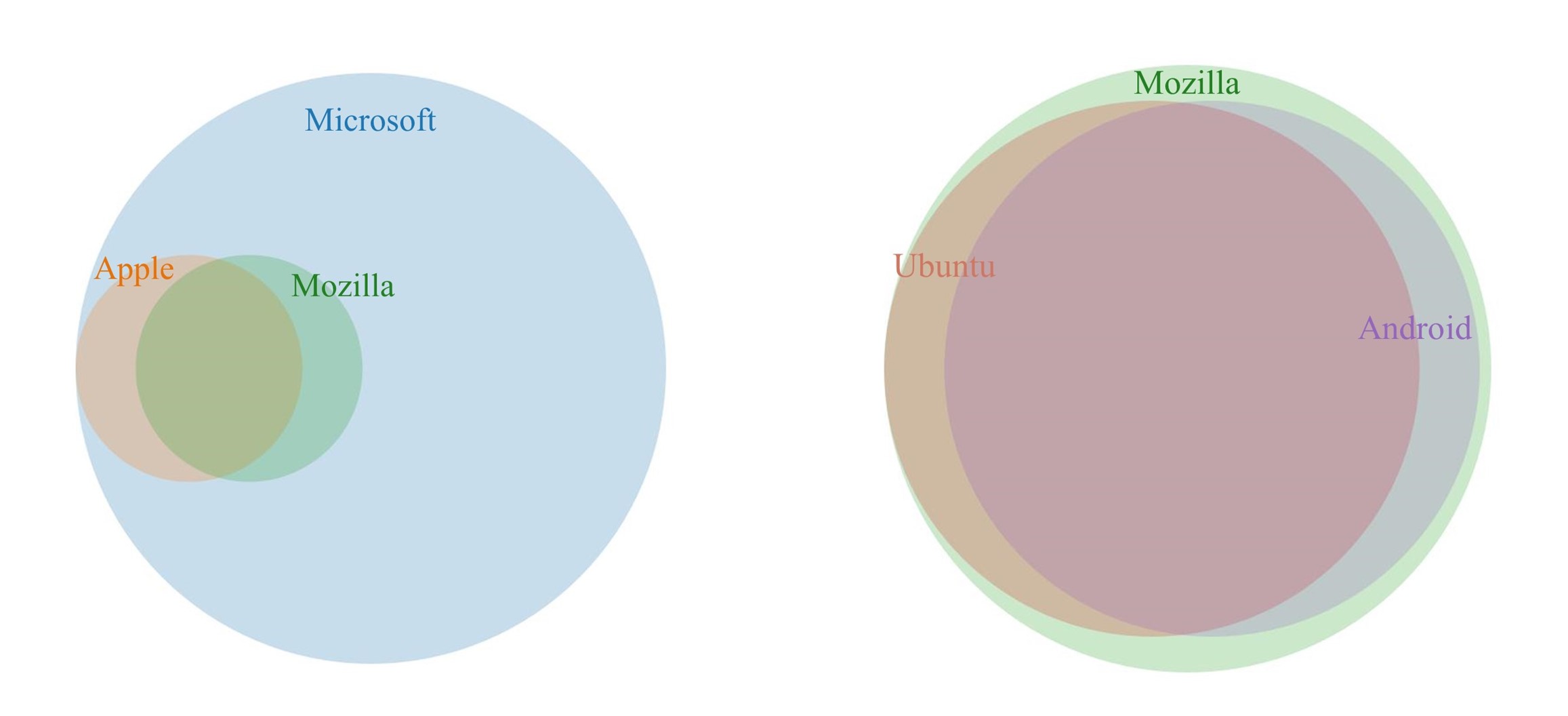}}
\caption{Government certificates in root stores.}
\label{fig:govcert}
\end{figure}

\begin{figure}[htbp]
\centerline{\includegraphics[width = 0.5 \textwidth]{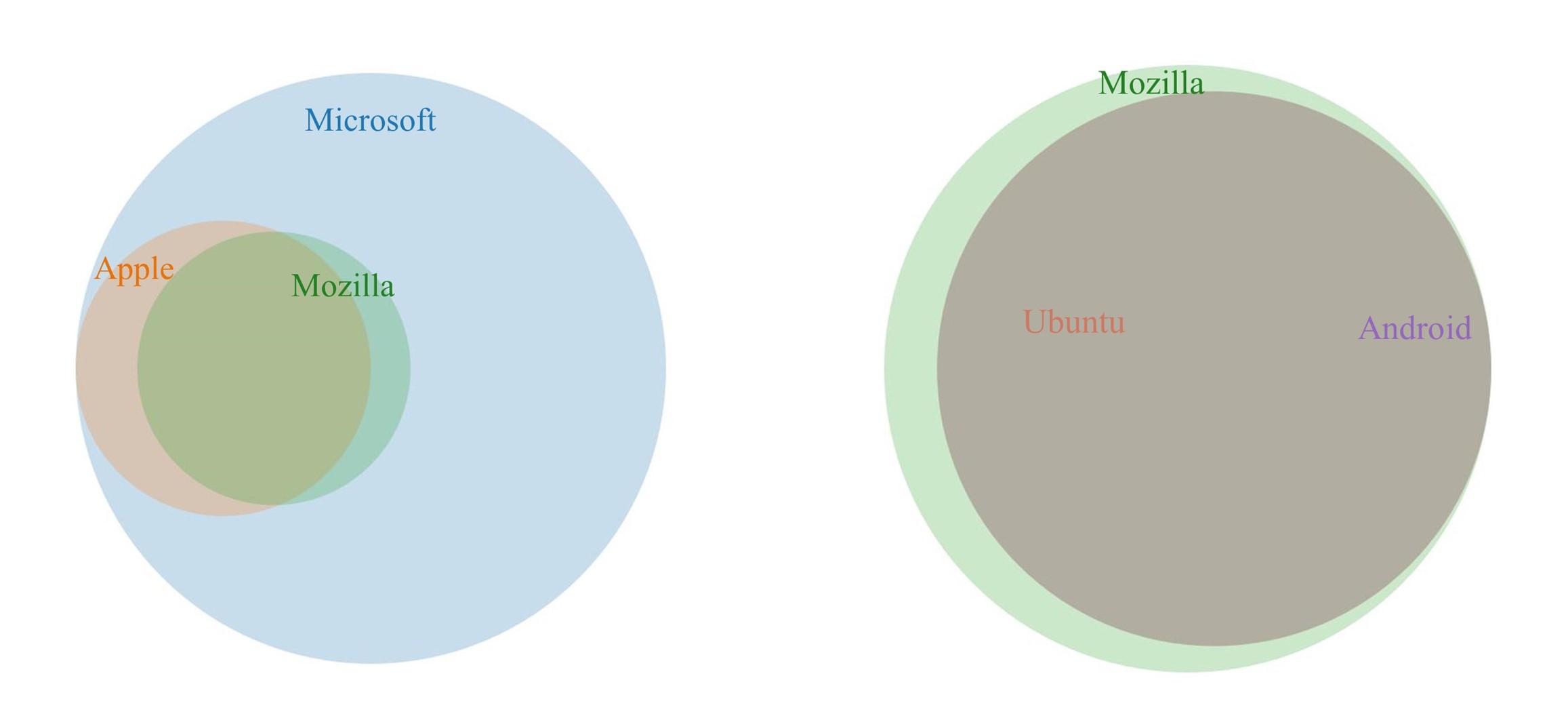}}
\caption{Governments trusted by vendors.}
\label{fig:govorg}
\end{figure}

Counting the number of government-issued root certificates, we find that Microsoft has 61, Apple has 11, NSS-based stores have 7, and Mozilla has 9. The overlap of such certificates between vendors is shown in Fig.~\ref{fig:govcert}, and of government organizations in Fig.~\ref{fig:govorg}. Note that even though Mozilla appears to have some exclusively-trusted government certificates in Fig.~\ref{fig:govcert}, these are issued by governments that are trusted by the other vendors (notice how Mozilla's set is fully enclosed within the others in Fig.~\ref{fig:govorg}). We found that 49 of the 61 government root certificates in Microsoft's store are exclusively trusted by Microsoft. Such certificates belong to the governments of Brazil, France, India, South Africa, Saudi Arabia, Uruguay, among others. We also found that Microsoft is the only vendor that trusts the U.S. government, but no government was found to be exclusively trusted by Mozilla, Apple, and the other NSS-based vendors. All root stores do contain certificates issued by the governments of Hong Kong, Spain, the Netherlands, and Turkey.

The above data accounts for certificates that have the words \emph{Government}, \emph{National}, \emph{Federal}, \emph{GRCA},\footnote{Stands for Government Root Certification Authority.} and a government name in the \texttt{subject} field of the certificate. We manually verified that all resultant certificates do belong to governments. However, it is possible that we missed other government-controlled certificates. The are two reasons for that.

First, even if a certificate is not directly issued by a government, the issuing organization could be partially owned by the government. For example, Deutsche Telekom is $\sim$32\% owned by the German government, and has at least two root certificates in all root stores.\footnote{\url{https://wiki.mozilla.org/CA:GovernmentCAs}} This is not counted in Fig.~\ref{fig:govorg}.

Second, certificate cross-signing leads to the indirect expansion of trust. However, CT-logs can alleviate this concern. Hiller \etal~\cite{hiller2020boon} highlight that certificates signed by the US government's Federal PKI (FPKI) organizations were being trusted on Mozilla products despite the absence of FPKI certificates from Mozilla's root store. Data on government certificates in root stores alone is thus inadequate to identify the full extent of trusted governments, and the impact of cross-signatures on expanding the trust of government roots requires further investigation. 

For the above two reasons, the numbers on government root certificates we report herein reflect a lower bound on the extent of government control over the vendors' root stores.

\section{Closing Remarks}

We find that the overwhelming majority of leaf certificates extracted from Censys are trusted by all vendors. However, we argue that the disparities across vendors' root stores remain concerning. While some degree of root-store differences is expected because vendors and device manufacturers use their own certificates to support their platforms (\eg code signing for software updates, including browser extensions and smartphone apps), the differences we noticed are much deeper than their exclusively-trusted certificates. What are the consequences of Microsoft choosing to trust that many government organizations, which others have not? How does a user access the services of one of these governments using another browser where said government has no root certificate? The disparities go beyond the root certificates themselves; the lack of transparency of some vendors, and the lack of unity in root-store inclusion policies result in disproportionate susceptibility to trust-abuse.

In our personal view, government certificates in root stores are dangerous, and hardly justifiable. We believe that a government organization should simply request a certificate for its domains from a trusted CA, just like any organization. The issuing CA has no control over the private key of the entity requesting the certificate. Governments around the world attempt to spy on citizens time and again through TLS interceptions and ISP coercions. (In 2019, the government of Kazakhstan forced users to install government-owned root certificates ~\cite{raman2020investigating}.) While a non-government-owned CA can still be subjected to coercion to act in the government's interest~\cite{clark2021trust} (\eg Dark Matter~\cite{TCDarkmatter} and StartCom~\cite{startcom}), we believe it still raises the bar because a CA typically undergoes trust-approval tracks that often include gaining popularity, building reputation, complying with audits, and committing to various forms of corporate transparency.

Research focus on root stores started to pick-up momentum recently. Ma \etal~\cite{ma2021tracing} analyzed several root stores and their inclusion policies, focusing more on root-store changes over time (a longitudinal view). The authors also identified questionable practices in root store management, such as unjustified trust of controversial CAs/organizations. 
Korzhitskii \etal~\cite{korzhitskii2020characterizing} looked at the list of root CAs allowed to append certificates to CT logs (CT root stores). Zhang \etal~\cite{zhang2021rusted} investigated root stores from a large number of volunteering users, and compared them with the expected root stores that they should have on their systems. Such increased attention from the research community is promising. 

Finally, it is worth noting that root stores evolve rapidly. Over the course of our data collection and analysis (few months), some root stores did get updated. Unfortunately, disparities in all the aspects we shed light on remain in place: inclusion policies, delivery methods, root certificates, trust context, and trusted government organizations. We hope this position paper musters wider involvement from the community in discussing the path forward to allow for greater transparency about vetting practices, and wider community involvement in trust-related decision making.

\bibliographystyle{ACM-Reference-Format}
\bibliography{references}

\end{document}